\documentclass[journal]{new-aiaa}
\usepackage[utf8]{inputenc}
\usepackage{textcomp}
\usepackage{bm}
\usepackage{multirow}
\usepackage{subcaption}
\usepackage{caption}
\usepackage{graphicx}
\usepackage{amsmath}
\usepackage[version=4]{mhchem}
\usepackage{siunitx}
\usepackage{longtable,tabularx}
\usepackage[dvipsnames]{xcolor}
\usepackage{tikz}
\usetikzlibrary{calc,patterns,angles,quotes,positioning,shapes,shadows,arrows}
\tikzstyle{input} = [rectangle, rounded corners, minimum width=0.1cm, minimum height=1cm, text centered, text width=2cm, draw=black, fill=yellow!30]
\tikzstyle{algorithm} = [rectangle, rounded corners, minimum width=0.1cm, minimum height=1cm, text centered, text width=2cm, draw=black, fill=red!30]
\tikzstyle{output} = [rectangle, rounded corners, minimum width=0.1cm, minimum height=1cm, text centered, text width=2cm, draw=black, fill=green!30]
\tikzstyle{define} = [rectangle, rounded corners, minimum width=0.1cm, minimum height=1cm, text centered, text width=2cm, draw=black, fill=blue!30]
\tikzstyle{invisible} = [rectangle, rounded corners, minimum width=0.1cm, minimum height=1cm, text centered, text width=2cm]
\tikzstyle{arrow} = [thick, ->,>=stealth]

\title{Cooperative Navigation Using Pairwise Communication with Ranging and Magnetic Anomaly Measurements}

\author{Chizhao Yang\footnote{Ph.D. Student, Department of Mechanical and Aerospace Engineering, AIAA Student Member (Corresponding Author).}, Jared Strader\footnote{Ph.D. Student, Department of Mechanical and Aerospace Engineering. (Equal contribution with Corresponding Author)} and Yu Gu\footnote{Associate Professor, Department of Mechanical and Aerospace Engineering, AIAA Senior Member.}}
\affil{West Virginia University, Morgantown, WV, 26506, United States}
\author{Aaron Canciani\footnote{Assistant Professor, Department of Electrical Engineering.}}
\affil{Air Force Institute of Technology, Wright-Patterson AFB, OH, 45433, United States}
\author{Kevin Brink\footnote{Senior Research Engineer, Air Force Research Laboratory, AIAA Senior Member.}}
\affil{Air Force Research Laboratory, Eglin AFB, FL, 32547, United States}

\begin{document}

\footnotetext{Presented as Paper 2018-1595 at the 2018 AIAA Guidance, Navigation, and Control Conference, Kissimmee, FL, January 8–12, 2018; received 2 August 2019; revision received 13 May 2020; accepted for publication 14 May 2020; published online 3 June 2020.}

\maketitle

\begin{abstract}
The problem of cooperative localization for a small group of unmanned aerial vehicles (UAVs) in a Global Navigation Satellite System-denied environment is addressed in this paper. 
The presented approach contains two sequential steps: first, an algorithm called cooperative ranging localization, formulated as an extended Kalman filter, estimates each UAV’s relative pose inside the group using intervehicle ranging measurements; second, an algorithm
named cooperative magnetic localization, formulated as a particle filter, estimates each UAV’s global pose through matching the group’s magnetic anomaly measurements to a given magnetic anomaly map. 
In this study, each UAV is assumed to only perform a ranging measurement and data exchange with one other UAV at any point in time.
A simulator is developed to evaluate the algorithms with magnetic anomaly maps acquired from airborne geophysical survey. 
The simulation results show that the average estimated position error of a group of 16 UAVs is approximately 20 m after flying about 180 km in 1 h. 
Sensitivity analysis shows that the algorithms can tolerate large variations of velocity, yaw rate, and magnetic anomaly measurement noises. Additionally, the UAV group shows improved position estimation robustness with both high- and low-resolution maps as more UAVs are added into the group.
\end{abstract}

\section{Introduction}\label{s:introduction}
\lettrine{C}{ooperative} multiple Unmanned Aerial Vehicle (UAV) systems have become increasingly popular due to the wide range of applications they are supporting, such as surveillance, search and rescue, and mapping. 
Compared to a single UAV, a group of cooperative vehicles may provide several navigational benefits, such as reduced dead-reckoning error, tolerance against individual vehicle or sensor failures, distribution of sensors across a larger spatial area, and shared observations (e.g., landmark). 
In this paper, we present a cooperative UAV navigation algorithm to achieve accurate estimates of vehicles' position and orientation with respect to each other inside a group (i.e., relative pose) as well as to the geographic coordinate system (i.e., global pose). 
The information required by our algorithm only includes each vehicle’s velocity and yaw rate measurements, pairwise inter-vehicle ranging measurements, magnetic anomaly measurements, and a prior magnetic anomaly map.

Several approaches exist for cooperative localization that aim to estimate the relative poses inside a robot or UAV group using particle filters \cite{fox2000probabilistic}, maximum likelihood estimation \cite{howard2002localization}, maximum a posteriori estimation \cite{nerurkar2009distributed}, covariance intersection \cite{arambel2001covariance, ouimet2018cooperative}, split covariance intersection \cite{li2013cooperative}, least squares \cite{strader2016cooperative}, and Kalman filters \cite{mourikis2006performance, roumeliotis2002distributed, vetrella2017satellite}, to name a few. 
In these approaches, on-board sensory information (such as from Inertial Measurement Units, or IMUs) and relative observations (i.e., one vehicle detects and identifies another one and measures some relative quantity through its exteroceptive sensors, such as camera or light detection and ranging (LIDAR)) are often used jointly. 
To improve the reliability of relative observations, many approaches rely on both inter-vehicle ranging and bearing measurements \cite{fox2000probabilistic, kurazume1994cooperative}. 
Bearing-only measurements have been tested and proven capable of estimating the pose of two nearby vehicles (about 10 meters apart) \cite{martinelli2005multi}. 
However, bearing measurements often have limited range. 
Rapid and precise distance measurements at long range are available using a coherent laser ranging system presented in \cite{coddington2009rapid} or radio based systems \cite{getting1993perspective}. 
One UAV equipped with one ranging sensor, however, usually can only perform the ranging measurement with one other UAV in a long distance at each point in time \cite{coddington2009rapid}. 
In order to be suitable for the limitations of current communication and ranging technology, in this paper, each vehicle is constrained to communicate and perform ranging measurements with one other UAV at each point in time. 
This configuration is referred to as pairwise communication in this paper (see details in Section \ref{s:communication}).

To estimate the UAV group’s global pose, at least one UAV in the group should be able to localize itself with respect to global references. 
Traditionally, Global Navigation Satellite System (GNSS) and IMUs are integrated to provide vehicles’ global pose estimates \cite{wendel2006integrated}. 
However, GNSS is not always available or reliable—due to reasons such as signal blockages, multipath reflection, and jamming. 
In a GNSS-denied environment, alternative information sources, such as visual information from cameras \cite{chowdhary2013gps, scaramuzza2014vision, weiss2011monocular, ahn2015autonomous, magree2016factored} or ranging and bearing information from LIDAR \cite{bachrach2011range, tang2015lidar, gu2018robot, ohi2018design}, are often applied to regulate the localization error growth. 
Both camera-based and LIDAR-based positioning approaches are performed by matching similar features in two or more images or scans. 
Therefore, these approaches may not be suitable for vehicle operations over extended low feature or repetitive pattern surfaces, such as sea or desert.

The Earth’s magnetic anomaly information offers another alternative data source for estimating vehicles’ global positions \cite{mcelhinny1998magnetic, canciani2016absolute}. 
The Earth’s magnetic anomalies present the high spatial frequency content of the Earth magnetic field. 
Additionally, the magnetic anomaly information has been measured for most regions in the world \cite{canciani2016absolute}. 
Even in an indoor environment, the magnetic field can be mapped for supporting vehicles' navigation systems \cite{brzozowski2016concept, brzozowski2017magnetic}.
Another reason for using magnetic anomaly information in localization is due to the technological maturity of magnetometers. 
For instance, optically pumped cesium magnetometers, which can achieve an accuracy of 0.1 nanoTesla(nT), have been used to create magnetic anomaly maps in geological surveys \cite{grosz2017high, hood2007history}. 

Several research groups have performed single vehicle navigation studies using magnetic anomalies as the primary source of information \cite{canciani2016absolute, li2012feasible, vallivaara2011magnetic}. 
For example, Canciani et al. developed a navigation filter through a fingerprint matching method to successfully estimate a single UAV’s global pose using the Earth’s magnetic anomaly map and a navigation grade-INS (inertial navigation system) \cite{canciani2016absolute}. 
However, these magnetic anomaly-based navigation approaches were performed at a low altitude. 
As altitude increases, the spatial frequency content of a magnetic anomaly field decreases \cite{blakely1996potential}, which directly affects navigation performance. 
In Ref. \cite{canciani2017improved}, the magnetic anomaly navigation accuracy was improved through a particle filter based cooperative navigation framework. 

In this paper, we present a cooperative localization method for a group of UAVs using inter-vehicle ranging and magnetic anomaly measurements in a GNSS-denied environment. 
Through this method, each UAV in the group can accurately estimate its global pose by matching all UAVs’ magnetic anomaly measurements to the given magnetic anomaly map. 
The geometric structure of the group at each time step is defined by the relative poses inside the group, which are estimated using pairwise ranging information. 
Compared with the approach presented in the Ref. \cite{canciani2017improved}, we are able to estimate the global pose with lower computation requirements due to the significantly less number of states used in the particle filter. 
This would in turn allow the application of this cooperative navigation algorithm on a larger UAV group. 
Early results of this research were presented in \cite{yang2018cooperative}.
However, the previous analysis was not adequate for understanding the performance in more realistic scenarios.
Additionally, the group size was restricted to an even number of UAVs in \cite{yang2018cooperative}.
In this paper, the simulator is expanded to support the UAV control, which relies on the estimated poses for feedback.
The magnetic anomaly map used in this paper is obtained from the United States Geological Survey to cover a larger area \cite{usgs2014magnetic}, which is more realistic than the simulated map in \cite{yang2018cooperative}.
Meanwhile, in this paper, the performance and robustness of our methods under different sensor noise assumptions and map resolutions are reported.
The method is additionally expanded to support odd numbers of UAVs in the group.

The remaining sections of the paper proceed as follows. 
Section \ref{s:problem_statement} provides a problem statement with the associated assumptions. 
In Section \ref{s:technical_approach}, methods are presented for solving the relative localization problem as well as for performing cooperative magnetic localization. 
Section \ref{s:simulator} presents the simulator and the simulation configurations. 
Then, in Section \ref{s:results}, simulation results are presented and analyzed. 
Section \ref{s:conclusion} concludes the contributions of this paper and identifies the current limitations and future research directions.

\section{Problem Statement}\label{s:problem_statement}

In this study, the case is considered where a group of $N$ UAVs is entering a GNSS-denied environment. 
The size of $N$ is appropriate for a typical UAV mission (e.g., between 2 and 20). 
The main objective is to achieve accurate and robust global localization for all UAVs in the group.

All UAVs are assumed to be flying at the same altitude, which means that this study only deals with the navigation problem in 2D. 
The initial pose in the global frame for each UAV is assumed to be known with a small uncertainty (e.g., provided by GPS) before entering the GNSS-denied environment.

Each UAV is assumed to be equipped with radios that enable communication with all other UAVs to exchange sensor measurements, as well as to perform inter-vehicle ranging. 
As mentioned earlier, each UAV is restricted to communicate and perform ranging measurements with only one other UAV at any point in time, which is referred as to pairwise communication in this paper. 
Also, each UAV is assumed to be able to store and exchange information of the other UAVs. 

Each UAV also performs point measurements of the magnetic anomaly field at each time step, which can be exchanged through the communication link. 
In addition, each UAV is assumed to be loaded with the same magnetic anomaly map. 
Meanwhile, each UAV is assumed to be able to measure or estimate its velocity and yaw rate through some methods, such as using Vehicle Dynamic Model (VDM) presented in \cite{khaghani2016autonomous} or Visual Odometry presented in \cite{strydom2014visual, caballero2009vision}. 
No additional information is used for UAV navigation in this study.

\section{Technical Approach}\label{s:technical_approach}

\subsection{Communication}\label{s:communication}
\begin{figure}[hbt!]
    \centering
    \includegraphics[width=.5\textwidth]{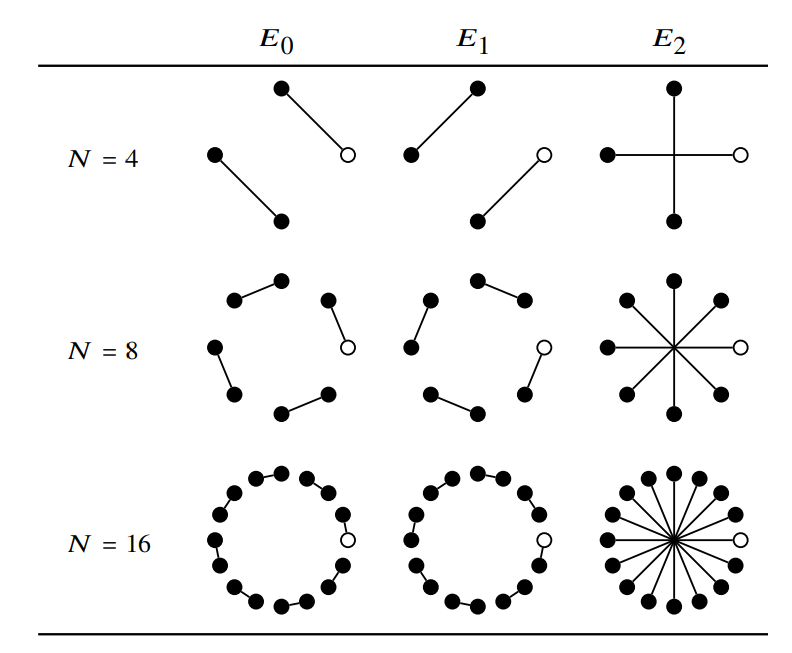}
    \caption{Pairwise communication graphs where the white nodes represent $\bm{v}_{\mathbf{1}}$ and the black nodes represent $\bm{v}_{\mathbf{2}}$, $\bm{v}_{\mathbf{3}}$, $\cdots$, $\bm{v}_{\bm{N}}$ in counterclockwise in each graph.}
    \label{fig:pairwise_communication}
\end{figure}

This subsection provides an overview of the proposed communication strategy and its limitations. 
Under pairwise communication as introduced in Section \ref{s:problem_statement}, multiple time steps are required in order for each UAV to exchange information with every other UAV.
Initially, in our previous work \cite{yang2018cooperative}, the communication strategy required an even number of UAVs. In this work, we expand for scenarios with an odd number of UAVs.
Consider the graph, $G=(V,E)$, representing the group of UAVs where $V$ and $E$ are the sets of vertices and edges in $G$, respectively. 
Let $V=\{v_1, v_2, \cdots, v_N\}$ such that each vertex corresponds to a single UAV. 
Since every vertex in $G$ is incident to exactly one edge in $E$, at least $\frac{(2L)!}{2^LL!}$ possibilities exist for $E$ where $N=2L \enspace (L \in \mathbb{Z}^+)$ \cite{west1996introduction}. 
To reduce the number of possibilities for $E$, only the edge sets given by $E_0=\{v_1v_2,v_3v_4, \cdots, v_{N-1}v_N\}$, $E_1=\{v_2v_3,v_4v_5,\cdots,v_Nv_1\}$, and $E_2=\{v_1v_{\frac{N}{2}+1},\ v_2v_{\frac{N}{2}+2},\ \cdots,\ v_\frac{N}{2}v_N\}$ are considered in this study. 
Note these edge sets produce all possible perfect matchings for $N = 4$; however, if $N > 4$, additional perfect matchings exist but are not selected as edge sets. 
Now, let $E$ be defined by
\begin{equation}\label{eq:E}
    E \in \{E_0,\ E_1,\ E_2\}
\end{equation}
where $E$ is used at each iteration of the cooperative localization algorithm introduced in Section \ref{s:cooperative_ranging_localization}.
Each of the edge sets produces an isomorphic graph; however, the isomorphisms are not label-preserving, so each edge set provides a different set of measurements. 
The graphs for $N=4,\ 8,$ and $16$ are presented in Fig. \ref{fig:pairwise_communication}.

\begin{figure}[hbt!]
    \centering
    \includegraphics[width=\textwidth]{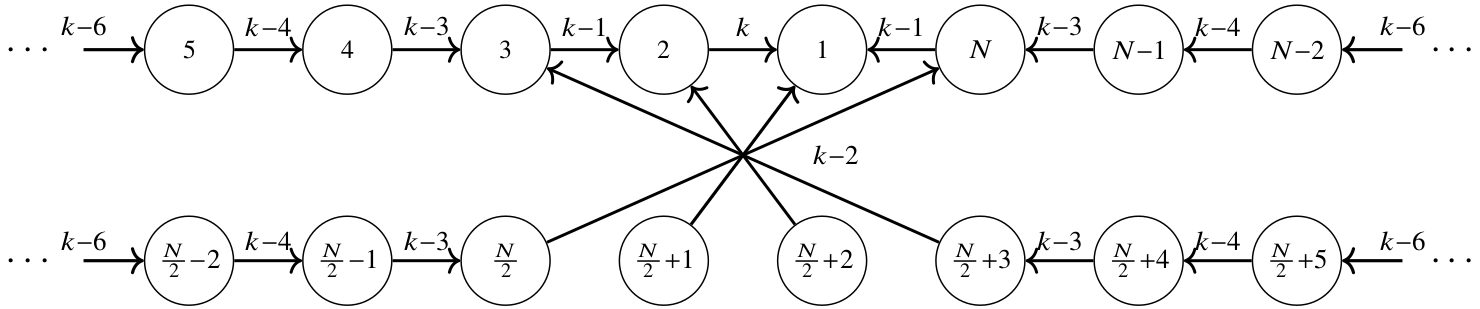}
    \caption{Propagation of information for pairwise communication by cycling between $\bm{E}_\mathbf{0}$ at $\bm{k}$, $\bm{E}_\mathbf{1}$ at $\bm{k}-\mathbf{1}$, and $\bm{E}_\mathbf{2}$ at $\bm{k}-\mathbf{2}$ repeatedly.}
    \label{fig:propagation_information}
\end{figure}
An important aspect of the communication strategy is the number of steps required to propagate a piece of information throughout the entire group of UAVs. 
In order to calculate the required number of steps, a graph can be constructed representing the propagation of information with respect to the first UAV. This is shown in Fig. \ref{fig:propagation_information} where the edges represent the connections made at each step and the vertices represent each source of information. 
Therefore, after $k$ discrete time steps, the length of the path from $v_i$ to $v_1$ is the number of steps required to propagate a piece of information from $v_i$ to $v_1$.

Now, consider the number of vertices added to the graph at each step prior to $k-2$. 
For $E_0$ and $E_1$, exactly four vertices are added to the graph separately. 
For $E_2$, each edge only provides a connection to an existing vertex, so the number of vertices remain the same. 
Thus, vertices are only added to the graph for $E_0$ and $E_1$ prior to $k-2$. 
For steps $k$ and $k-1$ exactly two vertices are added to the graph separately, and for steps $k-2$ exactly four vertices are added to the graph, as shown in Fig. \ref{fig:propagation_information}. 
Therefore, the number of vertices reached after $k$ discrete time steps is given by
\begin{equation}\label{eq:steps}
m=\begin{cases}
 & 2(k+1) \quad\quad \text{ if } k \leq 1 \\ 
 & 4k-4\left \lfloor \frac{k-2}{3}\right \rfloor\text{ otherwise } 
\end{cases}
\end{equation}
where $k\in\ \{0,1,2,\ \cdots\ \}$ and $\left\lfloor\ \right\rfloor$ is the floor function. 
Now, the number of steps required for one UAV to obtain the information of the rest of the group, denoted $s$, is bounded by $s\ \geq\left\lceil\frac{3}{8}N\ \right\rceil$ where $N$ is the number of UAVs in the group and $\left\lceil\ \right\rceil$ represents the ceiling function.

If the group size is an odd number (e.g., $N-1$), the bound for the $N-1$ UAV case is at most the same as for the $N$ UAV case. This is straightforward from the previous discussion. 
Consider the communication graph for $N$ UAVs (as depicted in Fig. \ref{fig:propagation_information}). 
Now, delete a vertex and the incident edges to form a group of $N-1$ UAVs, the resulting graph is still connected, and the number of steps required to propagate information remains the same. The resulting edge sets for an odd number of UAVs are obtained by deleting $v_N$ and all edges incident to $v_N$. The odd edge sets are then given by $E_0^{\text{odd}}=E_0^{\text{even}} \backslash v_{N-1}v_N$, $E_1^{\text{odd}}=E_1^{\text{even}} \backslash v_Nv_1$, and $ E_3^{\text{odd}}= E_3^{\text{even}} \backslash v_{\frac{N}{2}} v_N$.

\subsection{Cooperative Ranging Localization}\label{s:cooperative_ranging_localization}
The goal of the cooperative ranging localization algorithm is to to obtain a reliable estimate of the relative position of each UAV in the group for input to the cooperative magnetic localization algorithm introduced in Section \ref{s:cooperative_magnetic_loalization}. 
The problem is formulated as a state estimation problem with an EKF. 
In general, assuming additive noise, a discrete nonlinear dynamic system can be described by the state transition and observation models provided by
\begin{equation}\label{eq:ekf:dynamic}
    \mathbf{x}_k=\ \mathbf{f}(\mathbf{x}_{k-1},\ \mathbf{u}_k)+\ \mathbf{w}_k
\end{equation}
\begin{equation}\label{eq:ekf:observation}
    \mathbf{z}_k=\bm{h}(\mathbf{x}_k)+\mathbf{v}_k
\end{equation}
where $\mathbf{f}$ is the vector-valued discrete state prediction function, $\bm{h}$ is the vector-valued observation function, $\mathbf{x}$ is the state vector, $\mathbf{z}$ is the output vector, $\mathbf{u}$ is the measured control vector, $\mathbf{w}$ is the process noise vector, $\mathbf{v}$ is the measurement noise vector, and $k$ is the discrete time index. 
Both the process and measurement noises are assumed to be multivariate Gaussian white noises with covariances $\bm{Q}$ and $\bm{R}$, where $\mathbf{w}_k\sim N\left(0,\bm{Q}_k\right)$ and $\mathbf{v}_k\sim N\left(0,\bm{R}_k\right)$, respectively. 
The state vector in the EKF is given by
\begin{equation}\label{eq:ekf:state}
    \mathbf{x}=\left[\bm{\pi}^{\left(1\right)},\ \bm{\pi}^{\left(2\right)},\ \ldots,\ \bm{\pi}^{\left(N\right)}\right]^T
\end{equation}
where $\bm{\pi}^{\left(i\right)}=\left[x^{\left(i\right)},\ y^{\left(i\right)},\ \theta^{\left(i\right)}\right]^T$ is the pose of UAV $i$ in the global frame such that $i\in1,2,\cdots,N$ where $x$ and $y$ are the Cartesian coordinates for the position and $\theta$ is the heading.
The discrete time state transition function is given by $\mathbf{f}=\left[\mathbf{f}^{\left(1\right)},\ \mathbf{f}^{\left(2\right)},\ \ldots,\ \mathbf{f}^{\left(N\right)}\right]^T$ where $\bm{\pi}_k^{\left(i\right)}=\mathbf{f}^{\left(i\right)}\left(\bm{\pi}_{k-1}^{\left(i\right)},\mathbf{u}_k^{\left(i\right)}\right)$ is the discrete state transition function for UAV $i$ given by
\begin{equation}\label{eq:ekf:state_transition}
    \mathbf{f}^{(i)}(\bm{\pi}_{k-1}^{(i)},\mathbf{u}_k^{(i)})=
    \left[\begin{matrix}
    x_{k-1}^{(i)}\\
    y_{k-1}^{(i)}\\
    \theta_{k-1}^{(i)}\\\end{matrix}\right]
    +T_s 
    \left[\begin{matrix}v_k^{(i)}\cos{(\theta_{k-1}^{(i)}+ T_s\omega_k^{(i)})}\\
    v_k^{(i)}\sin(\theta_{k-1}^{(i)}+T_s \omega_k^{(i)})\\
    \omega_k^{(i)}\\\end{matrix}\right]
\end{equation}
where $v_k^{\left(i\right)}$ is the velocity for UAV $i$, $\omega_k^{\left(i\right)}$ is the yaw rate for UAV $i$ at time step $k$, and $T_s$ is the change in time between discrete time steps. 
Due to the constraints imposed by pairwise communication, the observation function varies between steps, which is given by
\begin{equation}
    \bm{h}\left(\mathbf{x}_{k|k-1}\right)=\sqrt{\left(x^{\left(i\right)}-x^{\left(j\right)}\right)^2+\left(y^{\left(i\right)}-y^{\left(j\right)}\right)^2}:\left(v_i,\ v_j\right)\in E_{k \; \left( \textsf{mod} \ 3\right)}
\end{equation}
where $\bm{h}$ is dependent on the edge set $E$ as described in Section \ref{s:communication}.
The measurement vector is given by $\mathbf{z}_k=\{d_{ij}:\left(v_i,v_j\right)\in E_{k \; \left(\textsf{mod} \ 3\right)}\}$ where $d_{ij}$ is the distance between UAV $i$ and UAV $j$ as measured by the ranging sensor. 
The standard EKF equations \cite{simon2006optimal} are used for the state prediction and update. 

As discussed previously, multiple steps are required for the UAVs to propagate information throughout the entire group. 
The number of required steps can be computed from Eq. \ref{eq:steps}. 
Therefore, the state estimate at any point in time is obtained by adding steps of dead-reckoning onto the state estimate obtained from most recent EKF update. 
This is given by
\begin{equation} \label{eq:ekf:final}
    \hat{\mathbf{x}}_k= \mathbf{x}_{k-s} + \Delta\mathbf{x}\bigr\rvert^k_{k-s}
\end{equation}
where $s$ is the number of steps since the most recent EKF update and $\Delta\mathbf{x}\bigr\rvert^j_i$ is the change in pose between time steps $i$ and $j$ from dead-reckoning. 
The relative position of each UAV with respect to other UAVs in the group can be derived through
\begin{equation}\label{eq:ekf:relative_position}
    \left[\begin{matrix}{\hat{x}}^{\left(i/j\right)}\\{\hat{y}}^{\left(i/j\right)}\\\end{matrix}\right]= \left[\begin{matrix}{\hat{x}}^{\left(i\right)}\\{\hat{y}}^{\left(i\right)}\end{matrix}\right]-\left[\begin{matrix}{\hat{x}}^{\left(j\right)}\\{\hat{y}}^{\left(j\right)}\end{matrix}\right]
\end{equation}
where $[ {\hat{x}}^{\left(i/j\right)},\, {\hat{y}}^{\left(i/j\right)} ]^T$ are the relative position coordinates for UAV $i$ with respect to UAV $j$, $[{\hat{x}}^{(i)},{\hat{y}}^{(i)}]^T$ and $[{\hat{x}}^{(j)},\ {\hat{y}}^{(j)}]^T$ are the position estimates of UAV $i$ and UAV $j$ from Eq. \ref{eq:ekf:final}, respectively.

Some limitations of the cooperative ranging localization should be mentioned in regard to observability. 
First, if the UAVs are traveling at same speed and direction with parallel trajectories, an infinite number of solutions exist for the relative position of each UAV. 
This phenomenon is discussed in detail in \cite{strader2016cooperative} and could potentially result in divergence of the EKF. 
This can be avoided by varying the velocity of each UAV and is discussed further in Section \ref{s:simulator}. 
Second, since ranging-only measurements are used for updating the poses, the EKF is only capable of preserving the pairwise distances. 
As a result, only the geometric structure of the group is maintained through the EKF. 
Thus, the geometric structure will rotate in the global frame with an angle $\gamma$, which is exactly the rotation error of the group if the geometric structure is known without error. 
This is a well-studied property described by Euclidean distance matrices \cite{gower1985properties, dokmanic2015euclidean}. 
The method for approximating $\gamma$ to reduce the rotation error and recover the global pose of the group is discussed in the following section.

\subsection{Cooperative Magnetic Localization}\label{s:cooperative_magnetic_loalization}
The goal of the cooperative magnetic localization algorithm is to bound the error growth of the global pose by matching the magnetic anomaly measurements on-board each UAV in the group to a prior magnetic anomaly map. 
This is achieved by leveraging the cooperative ranging localization solution to fix the relative position of each UAV (i.e., the group shape). 
The global position is maintained with a Bayes filter using a prediction-update framework. 
Since the magnetic anomaly map is highly non-linear, a particle filter is selected for solving this problem.

\begin{figure}[hbt!]
    \centering
    \begin{tikzpicture}
    	\node[draw,fill=gray,circle,inner sep=0.05cm,minimum width = 8,font=\scriptsize] (1) at (0,0) {};
    	\node[draw,fill=white,circle,inner sep=0.05cm,minimum width = 8,font=\scriptsize] (2) at (2.20,1.33) {};
    	\node[draw,fill=white,circle,inner sep=0.05cm,minimum width = 8,font=\scriptsize] (3) at (2.83,2.67) {};
    	\node[draw,fill=white,circle,inner sep=0.05cm,minimum width = 8,font=\scriptsize] (4) at (1.16,4) {};
    
    	\path[][thick,above, font=\scriptsize] (1) edge node {} (2);
    	\path[][thick,above, font=\scriptsize] (2) edge node {} (3);
    	\path[][thick,above, font=\scriptsize] (3) edge node {} (4);
    	\path[][thick,above, font=\scriptsize] (4) edge node {} (1);
    
    	\node[draw,fill=gray,circle,inner sep=0.05cm,minimum width = 8,font=\scriptsize] (11) at (0,0) {};
    	\node[draw,fill=black,circle,inner sep=0.05cm,minimum width = 8,font=\scriptsize] (22) at (1.54,2.06) {};
    	\node[draw,fill=black,circle,inner sep=0.05cm,minimum width = 8,font=\scriptsize] (33) at (1.63,3.53) {};
    	\node[draw,fill=black,circle,inner sep=0.05cm,minimum width = 8,font=\scriptsize] (44) at (-0.42,4.14) {};
    
    	\path[][thick,dashed,above, font=\scriptsize] (11) edge node {} (22);
    	\path[][thick,dashed,above, font=\scriptsize] (22) edge node {} (33);
    	\path[][thick,dashed,above, font=\scriptsize] (33) edge node {} (44);
    	\path[][thick,dashed,above, font=\scriptsize] (44) edge node {} (11);
    
    	\pic [draw, thick, angle radius=8mm, ->, "$\gamma$", angle eccentricity=1.5] {angle = 2--11--22};
    \end{tikzpicture}

    \caption{Example of rotation error of the group in relative position extracted from EKF for $\bm{N}=\mathbf{4}$.}
    \label{fig:relative_bearing_error}
\end{figure}
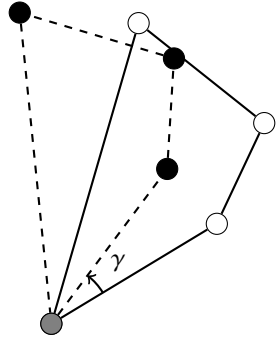

For reducing the rotation error of the group during flight, we define $\gamma$ as a state. 
An example of the rotation error of the group is presented in Fig. \ref{fig:relative_bearing_error} for the four-UAV case. 
In Fig. \ref{fig:relative_bearing_error}, the gray node is the position of UAV $j$, the black nodes connected with dashed lines are estimated relative positions of remaining UAVs with respect to UAV $j$ from the EKF, and the white nodes connected with solid lines are the true relative positions of remaining UAVs with respect to UAV $j$. 
The relative position estimates of each UAV with respect to the other UAVs in the group can be corrected based on $\gamma$ as,
\begin{equation}\label{eq:pf:relative_position}
    \left[\begin{matrix}{\grave{x}}^{\left(i/j\right)}\\{\grave{y}}^{\left(i/j\right)}\\\end{matrix}\right]=\left[\begin{matrix}\cos\gamma&-\sin\gamma\\\sin\gamma&\cos\gamma\\\end{matrix}\right]\left[\begin{matrix}{\hat{x}}^{\left(i/j\right)}\\{\hat{y}}^{\left(i/j\right)}\\\end{matrix}\right]
\end{equation}
where $[{\grave{x}}^{\left(i/j\right)}, {\grave{y}}^{\left(i/j\right)}]^T$ are the relative position coordinates for UAV $i$ with respect to UAV $j$ after applying the update utilizing $\gamma$, and $[ {\hat{x}}^{\left(i/j\right)},\, {\hat{y}}^{\left(i/j\right)} ]^T$ are the relative position coordinates for UAV $i$ with respect to UAV $j$ from the cooperative ranging localization.

An intuitive way to design the filter is to include all UAVs’ pose in the state vector. 
However, more states lead to more complexity and greater chance of overfitting. 
Therefore, the particle filter introduced in this paper only involves four states and is independent of group size, 
\begin{equation}
    \mathbf{p}=\left[x,y,\theta,\gamma\right]^T
\end{equation}
where $x$ and $y$ are the Cartesian coordinates for the global position of a UAV, $\theta$ is the heading of a UAV, and $\gamma$ is the rotation error of the group introduced previously. 
Note that each UAV runs a particle filter individually to estimate its own pose. 
Thus, the state transition model for each particle is given by
\begin{equation}
    \mathbf{p}_k=\ \mathbf{g}\left(\mathbf{p}_{k-1},\ \mathbf{u}_k^P\right)+\ \mathbf{\nu}_k
\end{equation}
where $\mathbf{g}$ is the vector-valued discrete state prediction function, $\mathbf{u}_k^P$ is the measured control vector of a UAV, $\mathbf{\nu}_k$ is the process noise vector, and $k$ is the discrete time index. 
Similar to Eq. \ref{eq:ekf:state_transition}, the vector-valued discrete state prediction function is given by
\begin{equation}
    \mathbf{g}\left(\mathbf{p}_{k-1},\ \mathbf{u}_k^P\right)=\left[\begin{matrix}\begin{matrix}x_{k-1}\\y_{k-1}\\\end{matrix}\\\begin{matrix}\theta_{k-1}\\\gamma_{k-1}\\\end{matrix}\\\end{matrix}\right]+T_s\left[\begin{matrix}\begin{matrix}v_k\cos{(\theta}_{k-1}+T_s \omega_k)\\v_k\sin{(\theta}_{k-1}+T_s \omega_k)\\\end{matrix}\\\begin{matrix}\omega_k\\0\\\end{matrix}\\\end{matrix}\right]
\end{equation}
where $v_k$ is the velocity for the UAV, $\omega_k$ is the yaw rate for the UAV, and $T_s$ is the change in time between discrete time steps. 
The state $\gamma$ is propagated by random walk. 
The observation model for each particle is given by
\begin{equation}\label{eq:pf:observation}
    \mathbf{y}_k=\bm{h}_M\left(\mathbf{p}_k,\ \mathbf{r}_k\right)+\mathbf{\eta}_k
\end{equation}
where $\mathbf{y}_k$ is the observation vector and $\mathbf{\eta}_k$ is the measurement noise vector.
The relative position $\mathbf{r}_k$ is calculated from Eq. \ref{eq:ekf:relative_position}. 
Therefore, for each UAV, the other UAVs’ predicted global positions can be extracted by adding the updated relative positions from Eq. \ref{eq:pf:relative_position} to the estimate from the particle filter, shown as
\begin{equation}\label{eq:pf:global_position_from_relative}
    \left[\begin{matrix}
      \grave{x}^{(i)} \\ \grave{y}^{(i)}
    \end{matrix}\right] = 
    \left[\begin{matrix}
      x^{(j)} \\ y^{(j)}
    \end{matrix}\right] +
    \left[\begin{matrix}
      \cos\gamma & -\sin\gamma \\ \sin\gamma & \cos\gamma
    \end{matrix}\right]
    \left[\begin{matrix}
      \hat{x}^{(i/j)} \\ \hat{y}^{(i/j)}
      \end{matrix}\right]
\end{equation}
where $[\grave{x}^{(i)} \grave{y}^{(i)}]^T$ is another UAV's predicted global position, $x^{(j)}, y^{(j)}$, and $\gamma$ are from the UAV's state vector $\mathbf{p}$, and $[\hat{x}^{(i/j)}, \hat{y}^{(i/j)}]^T$ is the relative position from $\mathbf{r}$.
The vector-valued observation function $\bm{h}_M$ is used to extract the predicted magnetic anomaly measurements from the given map based on each vehicles’ predicted positions.

The goal at each time step is to approximate the posterior distribution $p\left(\mathbf{p}_k\middle|\mathbf{y}_{1:k}\right)$ using a set of weighted particles. 
Consider the set of M particles given by
\begin{equation}
    \mathbf{P}_k=\left[\mathbf{p}_k^{\left(1\right)},\mathbf{p}_k^{\left(2\right)},\cdots,\mathbf{p}_k^{\left(M\right)}\right]
\end{equation}
where $\mathbf{p}_k^{\left(i\right)}=\left[x_k^{\left(i\right)},\ y_k^{\left(i\right)},\ \theta_k^{\left(i\right)},\gamma_k^{\left(i\right)}\ \ \right]^T$ is the state vector for the $i^{th}$ particle. 
The corresponding weights are given by
\begin{equation}
    \mathbf{W}_k=\left[w_k^{\left(1\right)},w_k^{\left(2\right)},\cdots,w_k^{\left(M\right)}\right]
\end{equation}
where $w_k^{\left(i\right)}$ is the weight for the $i^{th}$ particle. 
According to the observation function described in Eq. \ref{eq:pf:observation}, for each particle, there are $N$ different predicted observations corresponding to the $N$ UAVs’ predicted positions. 
In this study, the magnetic anomaly measurement noise is assumed as a Gaussian white noise \cite{canciani2016absolute}, and the likelihood function can be calculated as a Gaussian distribution. 
Meanwhile, the magnetic anomaly measurements from different UAVs are assumed to be conditionally independent. 
Therefore, for each particle, the likelihood function based on all observations at time $k$ is given by
\begin{equation}
    p\left(\mathbf{y}_k\middle|\mathbf{p}_k\right)=\prod_{j=1}^{N}{p\left(y_k^{\left(j\right)}\middle|\mathbf{p}_k\right)}
\end{equation}
where $p\left(y_k^{\left(j\right)}\middle|\mathbf{p}_k\right)=\frac{1}{\sqrt{2\pi\sigma_m^2}}\exp\left(-\frac{\left[y_k^{\left(j\right)}-t_k^{\left(j\right)}\right]^2}{2\sigma_m^2}\right)$ such that $y_k^{\left(j\right)}$ is the predicted observation on UAV $j$ based on the predicted state vector $\mathbf{p}_k$, $t_k^{\left(j\right)}$ is the magnetic anomaly measurement from UAV $j$'s on-board magnetometer, and $\sigma_m$ is the standard deviation of the magnetic anomaly measurement. 

The $i^{th}$ particle's weight is given by the likelihood function of the observations at time $k$ and its normalized weight in the previous time step, $w_k^{\left(i\right)}=p\left(\mathbf{y}_k\middle|\mathbf{p}_k^{\left(i\right)}\right){\widetilde{w}}_{k-1}^{\left(i\right)}$ where ${\widetilde{w}}_{k-1}^{\left(i\right)}$ is the normalized weight. 
The weights $\mathbf{W}_k$ are then normalized such that the sum of all weights equal one. 
The expectation of the state at time $k$ is given by 
\begin{equation}
    E[\mathbf{P}_k] \approx \sum_{i=1}^{M}\tilde{w}_k^{(i)}\mathbf{P}_k^{(i)}.
\end{equation}
Also, similar to the cooperative ranging localization method discussed earlier, the cooperative magnetic localization approach requires additional steps to gather all information in the group due to the pairwise communication restriction.

\section{Simulator}\label{s:simulator}
\begin{figure}[hbt!]
    \centering
    \begin{tikzpicture}[node distance=3cm]
    \node (shared) [input, text width=1.8cm] {Shared Information};
    \node (parse) [algorithm, right of=shared, xshift=-0.5cm] {Parse Data};
    \node (ranging) [algorithm, right of=parse, yshift=1.2cm, xshift=2.5cm] {Cooperative Ranging Localization};
    \node (magnetic) [algorithm, below of=ranging, yshift=1cm, xshift=0cm] {Cooperative Magnetic Localization};
    \node (delay) [invisible, draw=black, text width=2cm, below of=magnetic, yshift = 1cm] {Incremental Dead-reckoning};
    \node (add) [invisible, text width=0.8cm, minimum height = 0.1cm, minimum width = 0.1cm, draw=black, right of=magnetic, yshift = -1cm, xshift=-0.7cm] {Add};
    \node (estimate) [invisible, text width=0.8cm, right of=add, xshift = -1.5cm] {$\hat{\mathbf{x}}_{k}$};
    \node (measured) [input, left of=delay, xshift=-2.5cm, yshift=0.8cm, minimum height=2cm] {Measured Information};
    
    \draw[dashed] ($(shared.south west)+(-0.2, -0.2)$) rectangle ($(parse.north east)+(0.2, 0.5)$) node [left, xshift=-2.4cm, yshift=-0.2cm] {Communication};
    
    \draw [arrow] (shared) -- (parse);
    \draw [arrow] (parse) |- 
    node [invisible, anchor=south, xshift=2cm, yshift=-0.2cm, text width=6cm] {$\mathbf{RG}_{k-s,k-s+1,k-s+2}$, $\mathbf{V}_{k-s}$, $\mathbf{\Omega}_{k-s}$} 
    (ranging);
    \draw [arrow] ($(parse.east)+(0,-0.4)$)|- 
    node [invisible, anchor=south, xshift=1.8cm, yshift=-0.2cm, text width=3cm] {$\mathbf{M}_{k-s}$}
    ($(magnetic.west)+(0,0.4)$);
    \draw [arrow] ($(measured.east)+(0,0.8)$) |- 
    node [invisible, anchor=south, xshift=1.6cm, yshift=-0.2cm, text width=3cm] {$\mathbf{v}^{(i)}_{k-s}$, $\mathbf{\omega}^{(i)}_{k-s}$}
    ($(magnetic.west)+(0,-0.4)$);
    \draw [arrow] ($(measured.east)+(0,-0.8)$) --
    node [invisible, anchor=south, xshift=0cm, yshift=-0.2cm, text width=4cm] {$\mathbf{v}^{(i)}_{k-s,...,k}$, $\mathbf{\omega}^{(i)}_{k-s,...,k}$}
    (delay);
    \draw [arrow] (measured) -- (parse);
    \draw [arrow] (ranging) -- node [invisible, anchor=west, xshift=-0.6cm] {$\mathbf{r}_{k-s}$} (magnetic);
    \draw [arrow] (magnetic.east) -| node [invisible, anchor=south, yshift=-0.2cm] {$\hat{\mathbf{x}}_{k-s}$}
    (add.north);
    \draw [arrow] (delay.east) -| node [invisible, anchor=north] {$\Delta\mathbf{x}\bigr\rvert^{k}_{k-s}$}
    (add.south);
    \draw [arrow] (add) -- (estimate);
    \end{tikzpicture}
    \caption{The overview of the proposed algorithms applied in the $i^{th}$ UAV.}
    \label{fig:algorithms}
\end{figure}
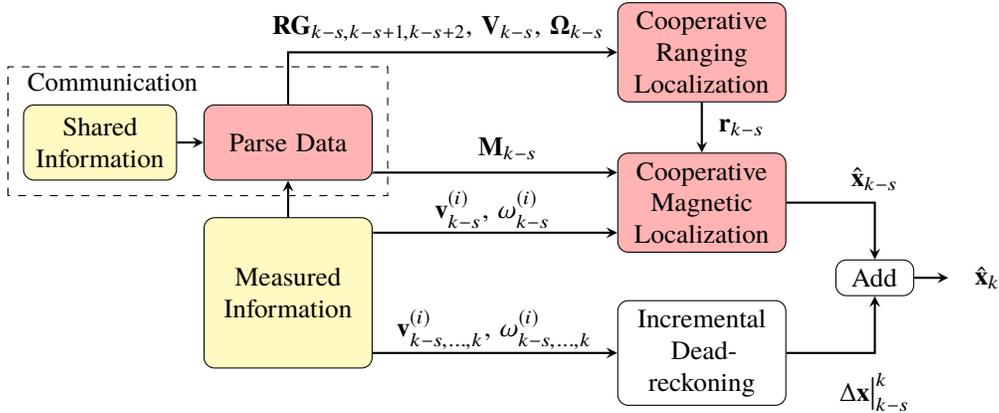

The cooperative ranging and magnetic localization algorithms are evaluated through a simulator, illustrated in Figs. \ref{fig:simulator} and \ref{fig:algorithms}.
Since all UAVs use the same simulation model, the pipeline of one UAV's simulator is presented in Fig. \ref{fig:simulator} where the purple blocks are user defined inputs and yellow blocks are measurements simulated from sensors and information shared from other UAVs through the communication links.
The pipeline of the proposed algorithms are explained in Fig. \ref{fig:algorithms} where $\mathbf{RG}_{k-s,k-s+1,k-s+2}$ are pairwise ranging measurements from the group at time steps $k-s, k-s+1$, and $k-s+2$, $\mathbf{V}_{k-s}$, $\mathbf{\Omega}_{k-s}$, and $\mathbf{M}_{k-s}$ present all UAVs' velocity, yaw rate, and magnetic anomaly measurements at time step $k-s$, respectively, $\mathbf{v}^{(i)}_{k-s}$ and $\mathbf{\omega}^{(i)}_{k-s}$ are the $i^{th}$ UAV's velocity and yaw rate measurements at time step $k-s$, respectively, $\mathbf{r}_{k-s}$ is the relative position estimates inside the group at time step $k-s$, $\hat{\mathbf{x}}_{k-s}$ is the global pose estimate at time $k-s$, and $\Delta\mathbf{x}\bigr\rvert^{k}_{k-s}$ is the change in global pose between time steps $k-s$ and $k$, as described in Eq. \ref{eq:ekf:final}.
Note that, in Fig. \ref{fig:algorithms}, the communication block is responsible for storing and parsing the historic data shared through the communication links with other UAVs and the data measured by sensors onboard the current UAV.
In detail, the data is stored in the shared information block as a set of packets, and the parse data block interprets the packets for input in the other parts of the system.
Each packet contains the data collected by all UAVs at time step $k$, denoted as $D_{k} = \{d^{(1)}_{k}, d^{(2)}_{k}, ..., d^{(N)}_{k}\}$ where $d^{(i)}_{k}$ is the data transferred from the $i^{th}$ UAV at time step $k$.
A packet is incomplete (denoted by $\grave{D}_k$) if the packet does not contain the data for all vehicles at time step $k$.
The parse data block is responsible for adding missing information to incomplete packets if data is available from packets received from other UAVs at a different time step.
Applying the communication scheme presented in Section \ref{s:communication}, the data transferred by each vehicle at time step $k$ is the time history given by $\mathbf{\overrightarrow{P}}_{k} = \{\grave{D}_{k}, \grave{D}_{k-1}, ..., D_{k-s}\}$ where all the packets are incomplete except for the packet containing data for time step $k-s$.
The packets for data recorded prior to time step $k-s$ are dropped as they are not used for localization.

\begin{figure}[hbt!]
    \centering
    \begin{tikzpicture}[node distance=3cm]
    \node (controller) [invisible, draw=black, minimum width=0.1cm, minimum height=2cm] {Controller};
    \node (model) [invisible, draw=black, above of=controller, yshift=-1cm] {Vehicle Model};
    \node (range) [input, text width=1.5cm, right of=controller, yshift=0cm, xshift=-0.9cm] {Pairwise Ranging Measurements};
    \node (magnetic) [input, text width=1.5cm, right of=range, xshift=-1.1cm] {Magnetic Anomaly Measurements};
    \node (shared) [input, text width=1.5cm, right of=magnetic, xshift=-1.1cm] {Shared Information};
    \node (map) [define, text width=1.5cm, right of=shared, xshift=-1.1cm] {Magnetic Anomaly Map};
    \node (proposed) [algorithm, right of=model, xshift=2.8cm, text width=2cm, minimum width=3cm, minimum height=1cm] {Proposed Algorithms (see Fig.\;\ref{fig:algorithms})};
    \node (traj) [define, left of=controller, xshift = 0.4cm, yshift=0.6cm] {Reference Trajectory};
    \node (vel) [define, left of=controller, xshift = 0.4cm, yshift=-0.5cm] {Reference Velocity};
    
    \draw [arrow] (controller) -- (model);
    \draw [arrow] (traj) -- ($(controller.west) + (0,0.5)$);
    \draw [arrow] (vel) -- ($(controller.west) + (0,-0.5)$);
    \draw [arrow] (magnetic) --(proposed);
    \draw [arrow] ($(range.north east)$) -- (proposed);
    \draw [arrow] (shared) --(proposed);
    \draw [arrow] (map) -- (proposed);
    \draw [arrow] (model) -- 
    node [anchor=east, xshift=1.6cm, yshift=0.2cm]{Velocity \& Yaw Rate}
    node [anchor=east, xshift=1.2cm, yshift=-0.2cm]{ Measurements}
    (proposed);
    \draw [arrow] (proposed.north) -- 
    ($(proposed.north) + (0,0.2)$) -| 
    ($(controller.west) + (-0.2,0.8)$) -- 
    node [invisible, anchor=east, xshift=-0.1cm, yshift=1.5cm]{Estimated Global Pose}
    ($(controller.west) + (0,0.8)$);
    \end{tikzpicture}
    \caption{The pipeline of the simulator for each UAV.}
    \label{fig:simulator}
\end{figure}
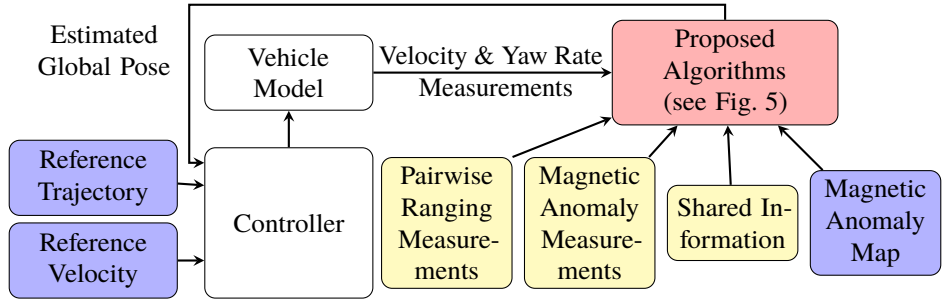

\begin{figure}[hbt!]
    \centering
    \includegraphics[width=.5\textwidth]{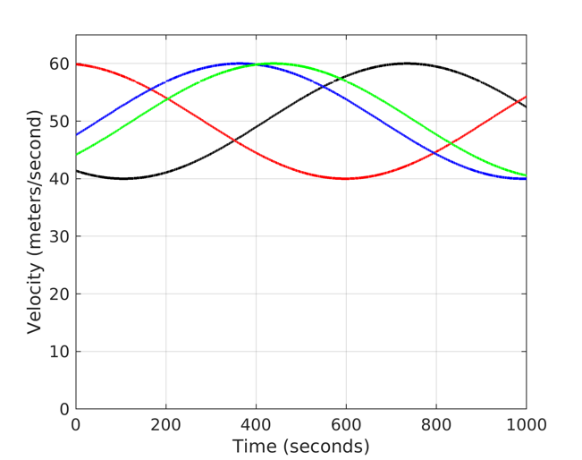}
    \caption{An example of the time-varying UAV reference velocities. Each UAV is presented with a different color.}
    \label{fig:example_velocity}
\end{figure}

In the simulator, a bicycle model \cite{corke2017robotics} is used as a simplified UAV model in the 2D plane. 
The trajectory of each UAV is controlled based on a reference trajectory (including position and velocity) and the estimated vehicle states. 
In this paper, all reference trajectories are set to be parallel to latitude lines in the geographic coordinate system, and new parallel reference trajectories, at a $1,000$ meters distance, are added to the simulation when the UAV group size increases. 
All UAVs are assumed to fly from west to east with the same initial longitude and small uncertainties on the initial positions, which are drawn from Gaussian distributions with zero mean and 1-meter standard deviation. 
Due to the observability requirement of the EKF, the reference velocity for each UAV is slowly varying following a sine function throughout the duration of the flight. 
The velocity offsets for each UAV are generated randomly so that all UAVs are traveling at different velocities. 
An example of the time-varying UAV reference velocities for the four-UAV case is shown in Fig. \ref{fig:example_velocity}.

\begin{figure}[hbt!]
    \centering
    \includegraphics[width=\textwidth]{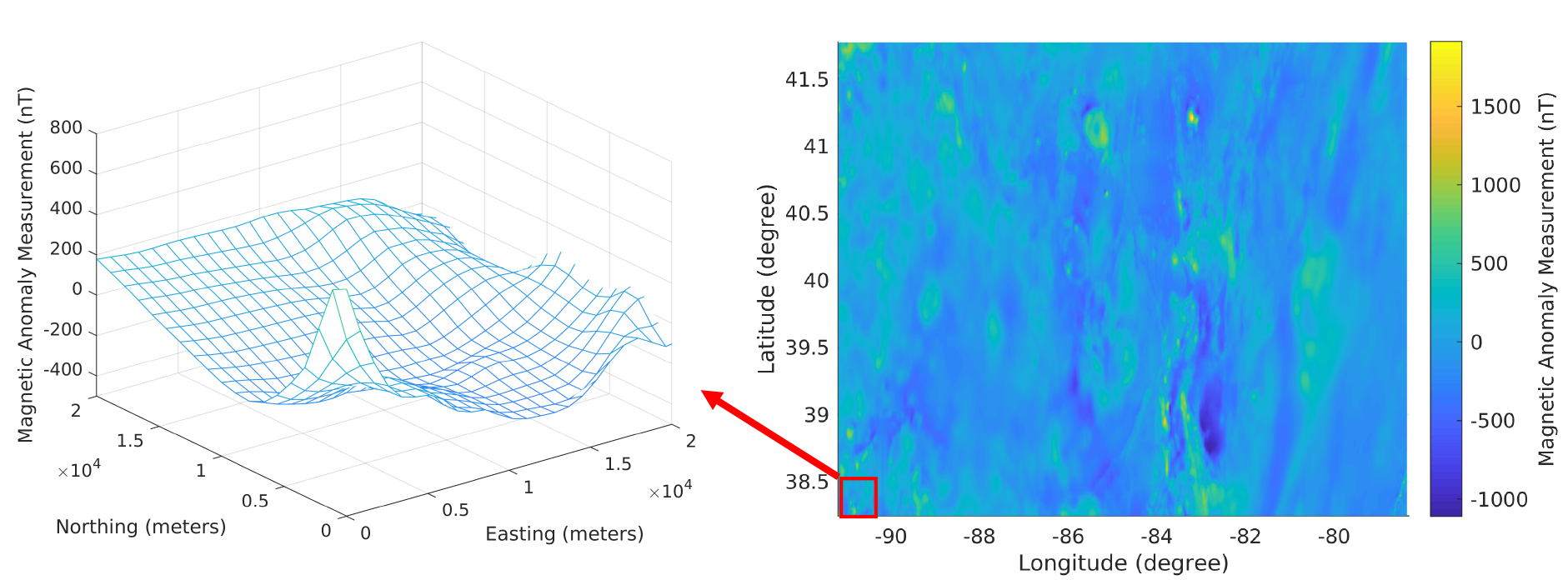}
    \caption{The magnetic anomaly map of the area around Columbus, Ohio, United State in geographic coordinate system (right). Left: The zoom-in part of the map.}
    \label{fig:magnetic_map}
\end{figure}

A controller steers the UAV based on the global pose estimates from the presented algorithms as feedback. 
Note that, in this paper, the velocity and yaw rate are measured at a frequency of 10 Hz, and the ranging measurements and magnetic anomaly measurements are obtained at a frequency of 5 Hz. 
The noise of the velocity measurement and yaw rate measurement are drawn from a Gaussian distribution with zero mean and standard deviation $\sigma_v$ for velocity and $\sigma_g$ for yaw rate. 
A turn-on bias $b_v$ for velocity and $b_g$ for yaw rate is added separately. 
Also, the noise of the ranging measurements and the magnetic anomaly measurements are drawn from a Gaussian distribution with zero mean and standard deviation $\sigma_r$ and $\sigma_m$, respectively.
The magnetic anomaly map utilized in this study, shown in Fig. \ref{fig:magnetic_map}, is obtained from the United States Geological Survey \cite{usgs2014magnetic} and contains the magnetic anomaly information at 305 meters altitude from the area around Columbus, Ohio, United States. 
The left part of the Fig. \ref{fig:magnetic_map} shows the zoom-in map of the area, which is marked as a red square on the geographic coordinate map. 
Note that, the unit in the zoom-in map (left) is converted from degree to meters along the east and north directions, and the left bottom corner of the geographic coordinate map is set as the original point in the zoom-in map.

\section{Results}\label{s:results}
\subsection{Case Studies}
The simulation case studies are performed for different group sizes ($N$ = 1, 3, 4, 7, 8, 15, and 16).
In each case, multiple Monte Carlo simulations are performed consisting of 200 trials each. 
The baseline simulation parameters are listed in Table \ref{table:baseline_parameter_noise} and Table \ref{table:baseline_parameter_trajectory}.
Based on the parameter setting of the reference velocity and flight duration, the trajectory length of each UAV is about 180 kilometers in each case.
\begin{table}[hbt!]
\caption{\label{table:baseline_parameter_noise} The baseline noise set in case studies.}
\centering
\begin{tabular}{cccccc}
\hline
$\sigma_r$ & $\sigma_m$ & $\sigma_v$ & $b_v$ & $\sigma_g$ & $b_g$ \\ \hline
1 meter & 10 nT & 0.3 m/s & $\sim N(0, 0.1\sigma_v)$ & 0.005 deg/s & $\sim N(0, 0.1\sigma_g)$ \\
\hline
\end{tabular}
\end{table}
\begin{table}[hbt!]
\caption{\label{table:baseline_parameter_trajectory} The baseline parameter set in case studies.}
\centering
\begin{tabular}{cccccc}
\hline
\multicolumn{4}{c}{Each UAV's reference velocity (sine function of time)} & \multirow{2}*{Number of particles} & \multirow{2}*{Filight duration} \\ \cline{1-4}
Amplitude & Baseline & Frequency & Phase && \\
\hline
10 m/s & 50 m/s & 0.05 s & rand*2*pi & 10,000 & 1 hour\\ 
\hline
\end{tabular}
\end{table}

Since the shape of the group is defined given the inter-vehicle distances, the inter-vehicle distance errors are used to analyze the quality of the group shape estimated by the cooperative ranging localization algorithm. 
Meanwhile, due to the limitation of the pairwise communication, the distances between some pairs of UAVs in the group are not measured by the ranging sensor at any point in time. 
The estimated distances are calculated based on the pose estimates from the cooperative ranging localization algorithm. 
The average distance error for each simulation is computed for measured pairs and not measured pairs separately.
Table \ref{table:distance_error} shows the means and standard deviations of these distance error for 200 Monte Carlo simulations.
\begin{table}[hbt!]
\caption{\label{table:distance_error} The means and standard deviations of the average distance error for 200 Monte Carlo simulations.}
    \centering
    \begin{tabular}{ccccc}
        \hline
         \multirow{2}*{Group size}& \multicolumn{2}{c}{Estimation with not measured edges} & \multicolumn{2}{c}{Estimation with measured edges} \\ \cline{2-5}
         & Mean (m) & Standard deviation (m) & Mean (m) & Standard deviation (m) \\ \hline
         3 & N/A & N/A & 0.6109 & 0.0061 \\
         4 & N/A & N/A & 0.5715 & 0.0050 \\
         7 & 0.8585 & 0.1112 & 0.6074 & 0.0034 \\
         8 & 0.7296 & 0.0299 & 0.5898 & 0.0030 \\
         15 & 1.0828 & 0.2373 & 0.6035 & 0.0022 \\
         16 & 0.8902 & 0.0692 & 0.5953 & 0.0021 \\
         \hline
    \end{tabular}
\end{table}

In Table \ref{table:distance_error}, most of the means of average distance errors are less than 1 m. 
The standard derivation of the ranging measurement noise is also set as 1 m in this case study. 
In other words, the cooperative ranging localization algorithm is able to maintain an accurate estimate of the inter-vehicle distances for both the measured and not measured pairs using pairwise communication.
This implies that the geometric structure (i.e., shape) of the group is maintained even in the case of missing range measurements. 
Interestingly, the distance errors of odd numbered groups are always larger than the nearest even numbered groups errors.
This is due to the fact that the total number of measured pairs in odd numbered groups is less than in the nearest even numbered groups as discussed in Section \ref{s:communication}, even though the nearest even number is larger than the odd number.
Note that, the distance errors are growing as the group size increases, which implies that the presented method is not scalable to a very large UAV group.

\begin{figure}[hbt!]
    \centering
    \begin{subfigure}{\columnwidth}
    \includegraphics[width=\textwidth]{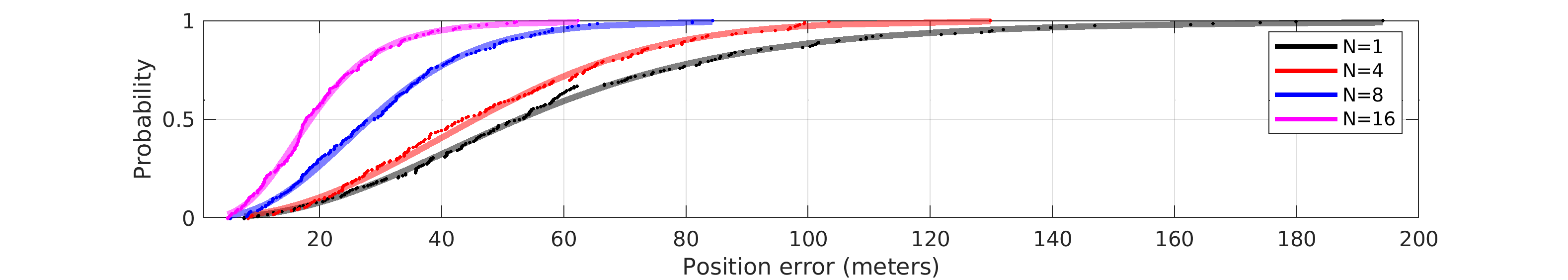}
    \caption{}
    \label{fig:group_size_even}
    \end{subfigure}
 
    \begin{subfigure}{\columnwidth}
    \includegraphics[width=\textwidth]{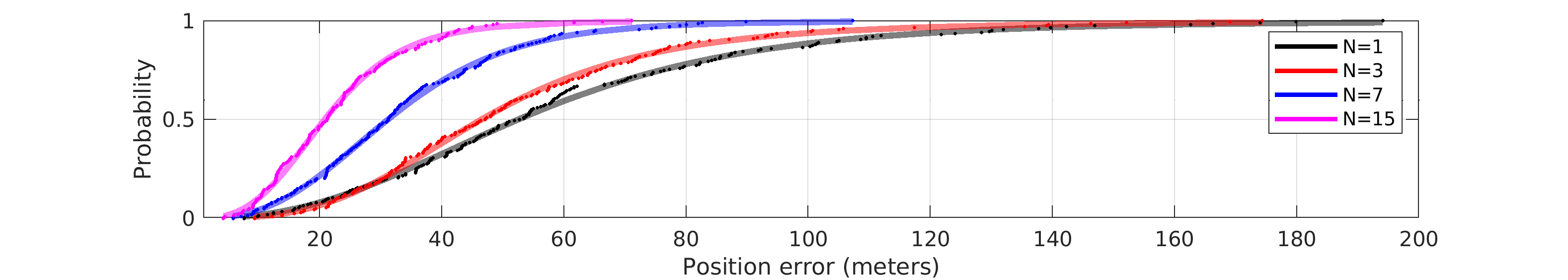}
    \caption{}
    \label{fig:group_size_odd}
    \end{subfigure}
    \caption{The CDF of average position error of one UAV in the group for 200 Monte Carlo simulations for different group sizes.}
    \label{fig:group_size}
\end{figure}

For evaluating the performance on global localization, the global position estimate errors of one UAV are reported since the algorithms running on each UAV are identical.
In other words, the average position error of one UAV in the group is computed for each simulation.
The Cumulative Distribution Function (CDF) and statistic data of average position error of one UAV in the group for 200 Monte Carlo simulations for different group sizes are shown in Fig. \ref{fig:group_size} and Table \ref{table:group_size} separately.
In Fig. \ref{fig:group_size}, the dots present the empirical CDF, the continues curves are approximated based on the dots.
The following CDF figures follow the same format as Fig. \ref{fig:group_size}.
The average position error is computed for each Monte Carlo simulation. 
The sizes of the group are $N=$ 1, 3, 4, 7, 8, 15 and 16.
Note that, for the $N=1$ case, since there is no relative information inside the UAV group, Eq. \ref{eq:pf:observation} in the particle filter is replaced by $\mathbf{y}_k=\bm{h}_M\left(\mathbf{p}_k\right)+\mathbf{\eta}_k$. 

\begin{table}[hbt!]
\caption{The statistical data of each simulation’s average position error of one UAV in the group for 200 Monte Carlo simulations.}
\centering
    \begin{tabular}{ccccc}
            \hline
             \multirow{2}*{Group size}& \multicolumn{2}{c}{Magnetic localization} & \multicolumn{2}{c}{Dead-reckoning} \\ \cline{2-5}
             & Mean (m) & Standard deviation (m) & Mean (m) & Standard deviation (m) \\ \hline
             1 & 59.4 & 35.3 & \multirow{7}*{741.3} & \multirow{7}*{552.9} \\
             3 & 52.3 & 28.7 & & \\
             4 & 47.8 & 23.6 & & \\
             7 & 33.9 & 17.6 & & \\
             8 & 30.3 & 15.2 & & \\
             15 & 22.6 & 11.5 & & \\
             16 & 20.2 & 10.3 & & \\
             \hline
    \end{tabular}
    \label{table:group_size}
\end{table}

The statistical data in Table \ref{table:group_size} shows substantial improvements in navigation performance using the presented navigation algorithm as compared to the dead-reckoning performance of each individual UAV. 
From the results shown in Fig. \ref{fig:group_size} and Table \ref{table:group_size}, we can see that the means and standard deviations of position error reduce when the group size increases for the simulated cases. 
This may due to increased spatial coverage of the UAV group and the increased number of magnetic anomaly measurements used for map matching.
Note that, in the simulated cases, the decrements of the position error are not proportional to the group size. 
The estimation errors may also be affected by the resolution of the magnetic anomaly map, the quality of sensors, and the error of the relative pose estimates.

\subsection{Sensitivity Analysis}
\begin{figure}[hbt!]
    \centering
    \includegraphics[width=\textwidth]{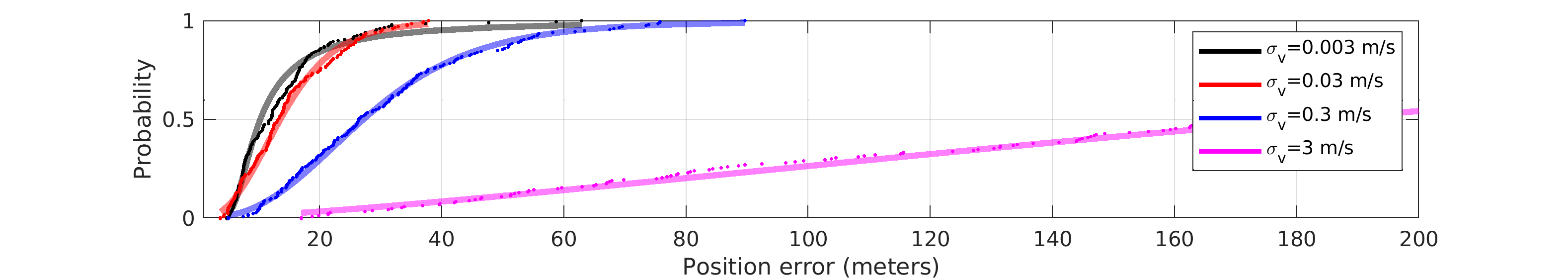}
    \caption{The CDF of each simulation’s average position error for one UAV in the group for 200 Monte Carlo simulations with varied velocity noise.}
    \label{fig:sigma_velcoity}
\end{figure}

\begin{figure}[hbt!]
    \centering
    \includegraphics[width=\textwidth]{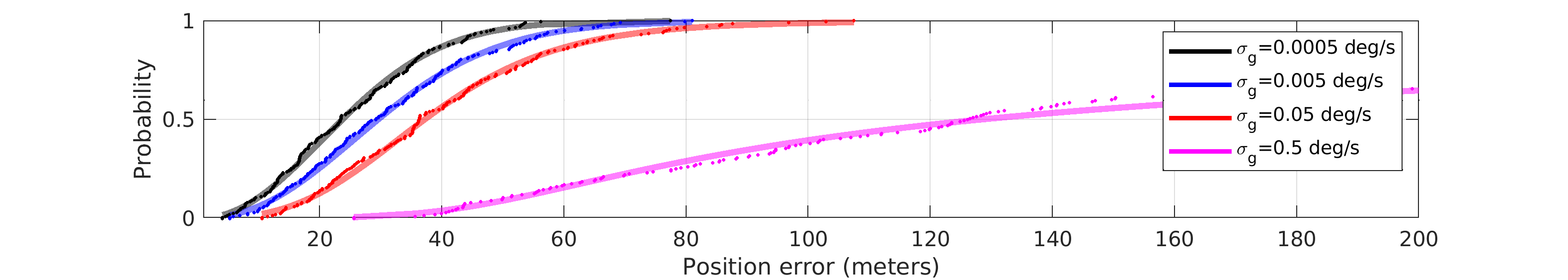}
    \caption{The CDF of each simulation’s average position error for one UAV in the group for 200 Monte Carlo simulations with varied yaw rate noise.}
    \label{fig:sigma_yawrate}
\end{figure}

\begin{figure}[hbt!]
    \centering
    \includegraphics[width=\textwidth]{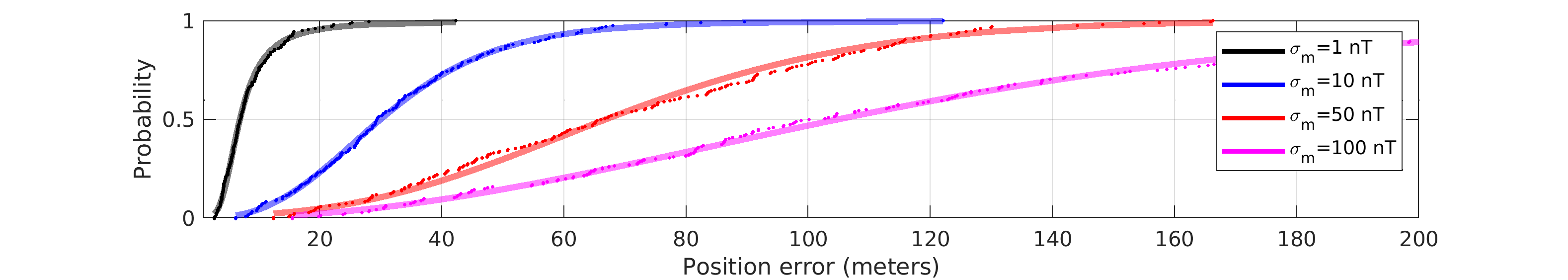}
    \caption{The CDF of each simulation’s average position error for one UAV in the group for 200 Monte Carlo simulations with varied magnetic anomaly measurement noise.}
    \label{fig:sigma_magnetic}
\end{figure}

Sensitivity analysis is performed to evaluate the effect of sensor quality as well as the resolution of the magnetic anomaly map on the presented algorithms. 
Simulations are first performed with a UAV group with $N$ = 8 and the baseline simulation parameters shown in Table \ref{table:baseline_parameter_noise} and Table \ref{table:baseline_parameter_trajectory}. 
In order to evaluate the impact of sensor noise on the algorithms presented in this paper, the simulations are performed while varying the velocity noise, yaw rate noise, and magnetic anomaly measurement noise separately. 
Similar as the case studies, multiple Monte Carlo simulations consisting of 200 trials each are performed in each case. 
Figures \ref{fig:sigma_velcoity}, \ref{fig:sigma_yawrate}, and \ref{fig:sigma_magnetic} present the CDF of each simulation’s average position error for one UAV in the group with varied velocity noise, yaw rate noise, and magnetic anomaly measurement noise, respectively. 
The blue lines are the baseline case in each figure. 
The sensitivity analysis results show that the developed cooperative navigation method can tolerate large variations of these sensor noises, with the exception of the cases with $\sigma_v=3\ m/s$  and $\sigma_g=0.5\ deg/s$. 

\begin{figure}[hbt!]
    \centering
    \includegraphics[width=\textwidth]{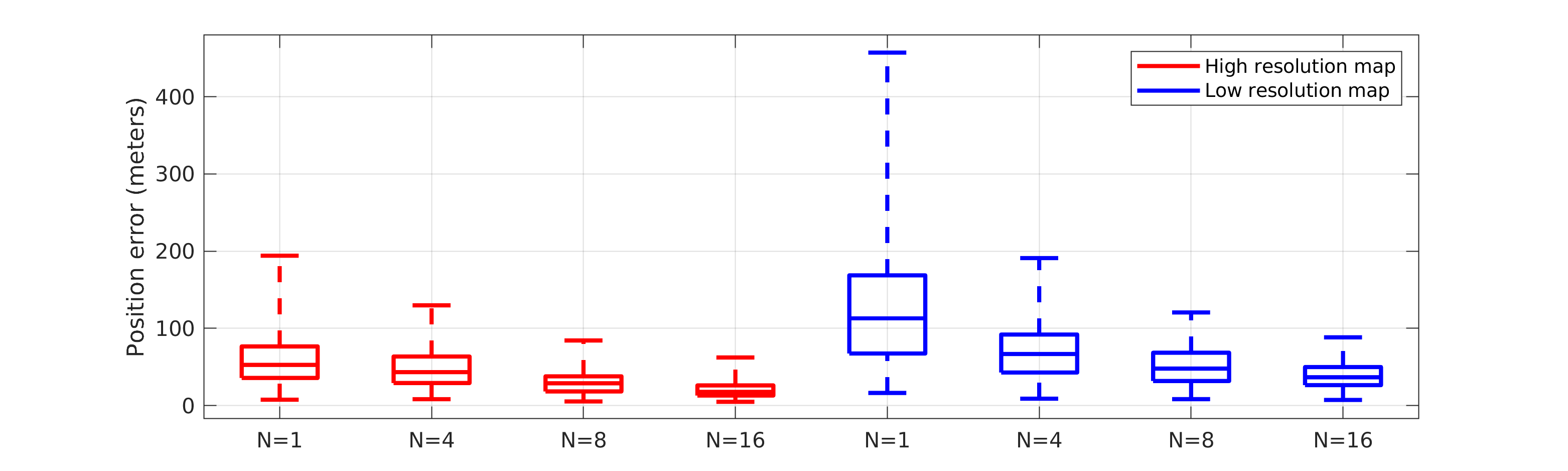}
    \caption{The boxplots of each simulation’s average position error for one UAV in the group with different group sizes using different resolution maps.}
    \label{fig:different_maps}
\end{figure}

In order to evaluate the algorithms in different resolution maps, another magnetic anomaly map is used in the simulations, which covers the same area as the map presented in Fig. \ref{fig:magnetic_map} but from a higher altitude, 3050 meters. 
As discussed in Section \ref{s:introduction}, when the altitude increase, the resolution of magnetic anomaly map decreases. 
In this case, the lower altitude map is called the high resolution map, and the higher altitude map is called the low resolution map. The algorithms are simulated for different group sizes ($N$ = 1, 4, 8, and 16) with different resolution maps. 
In each case, multiple Monte Carlo simulations are performed consisting of 200 trials each. 
Other parameters are set to be the same as ones shown in Table \ref{table:baseline_parameter_noise} and Table \ref{table:baseline_parameter_trajectory}. 
Figure \ref{fig:different_maps} shows the boxplots of each simulation’s average position error for one UAV in the group for 200 Monte Carlo simulations with different group sizes using different resolution maps. 
In each box, the five lines, from bottom to top, show the minimum, first quartile, median, third quartile, and maximum of the data separately. 

\section{Conclusion}\label{s:conclusion}
In this paper, a cooperative navigation algorithm was presented to estimate a group of UAVs’ global poses using inter-vehicle ranging and the Earth’s magnetic anomaly measurements. 
The approach contains two sequential steps:  cooperative ranging localization for relative navigation, formulated as an EKF, and cooperative magnetic localization utilizing a particle filter to estimate each UAV’s global pose. 
Furthermore, the presented algorithm is designed to perform using the pairwise communication, which is compatible with the current communication and ranging technology.

The simulation results show that the developed cooperative ranging localization algorithm is able to provide reasonable relative pose estimates. 
Utilizing the cooperative magnetic anomaly algorithm, each UAV can estimate its global pose with a high accuracy. 
Compared with a single UAV case, the cooperative UAV group can provide more accurate global pose estimates, as well as more robustness when using a lower resolution map. 
The results from the sensitive analysis show that the presented algorithm can tolerate large variations of velocity, yaw rate, and magnetic anomaly measurement noises. 

There are several limitations to the presented algorithms such as trajectory constraints. 
Specifically, the cooperative ranging localization algorithm will become unobservable when all UAVs are flying at the same speed along parallel trajectories.
These constraints will be relaxed in our future work through a tighter coupling between the cooperative ranging and magnetic localization steps. 
Additionally, the presented algorithm is not scalable to a very large UAV group, which will be a subject of our future work.
Also, in order to analyze the robustness of the presented algorithms for different geometries of the UAV group, various UAV trajectory settings will be applied for evaluation in the future. 
In addition, we will experiment the presented cooperative navigation algorithms with physical systems in the near future. 

\section*{Acknowledgments}
This research was supported in part by the US Air Force Research Laboratory under Award No. FA8651-16-1-0002, the Arlen G. and Louise Stone Swiger Doctorate Fellowship, and the West Virginia University Outstanding Merit Fellowship for Continuing Doctoral Students.

\bibliography{sample}

\end{document}